\documentclass[aps,prx,floatfix,singlecolumn,showpacs,preprintnumbers,amsmath,amssymb,superscriptaddress,notitlepage]{revtex4-1}%
\usepackage{graphicx}
\usepackage{dcolumn}
\usepackage{bm}

\begin{document}


\title{Crossover from Shear-Driven to Thermally Activated Drainage\\
 of Liquid-Infused Microscale Capillaries}
\author{Carlos E. Colosqui}
\email[]{carlos.colosqui@stonybrook.edu}
\affiliation{Department of Mechanical Engineering, Stony Brook University, Stony Brook, NY 11794, USA.}

\author{Jason S. Wexler}
\affiliation{Otherlab, San Francisco, CA 94110, USA.}
\affiliation{Department of Mechanical and Aerospace Engineering, Princeton University, Princeton, New Jersey 08544, USA.}
\author{Ying Liu}
\affiliation{Department of Mechanical and Aerospace Engineering, Princeton University, Princeton, New Jersey 08544, USA.}
\author{Howard A. Stone}
\email[]{hastone@princeton.edu}
\affiliation{Department of Mechanical and Aerospace Engineering, Princeton University, Princeton, New Jersey 08544, USA.}

%
%
%
%
%
%
\begin{abstract}
The shear-driven drainage of capillary grooves filled with viscous liquid is a dynamic wetting phenomenon relevant to numerous industrial processes and novel lubricant-infused surfaces.
Prior work has reported that a finite length $L_\infty$ of the capillary groove can remain indefinitely filled with liquid even when large shear stresses are applied.
The mechanism preventing full drainage is attributed to a balance between the shear-driven flow and a counterflow driven by capillary pressures caused by deformation of the free surface. 
%
%
In this work, we examine closely the approach to the final equilibrium length $L_\infty$ and report a crossover to a slow drainage regime that cannot be described by conventional dynamic models considering solely hydrodynamic and capillary forces. 
The slow drainage regime observed in experiments can be instead modeled by a kinetic equation describing a sequence of random thermally activated transitions between multiple metastable states caused by surface defects with nanoscale dimensions.
Our findings provide new insights on the critical role that natural or engineered surface roughness with nanoscale dimensions can play in the imbibition and drainage of capillaries and other dynamic wetting processes in microscale systems.  
\end{abstract}
%
%
%
\maketitle

%
%
%
\section{Introduction}
%
%
Dynamic wetting processes such as spreading, imbibition, and drainage are ubiquitous in  natural, agricultural, and industrial processes that are crucial to modern technology.
Engineering applications ranging from oil recovery and water treatment to microfluidics and bioanalytical systems have been enabled by a fundamental understanding of wetting that is embodied in mathematical descriptions such as the Young-Dupre, Young-Laplace, and Lucas-Washburn equations  \cite{de1985wetting,bonn2009wetting}. 
These classical wetting models are derived in the framework of continuum thermodynamics under the assumption of perfectly smooth and homogeneous surfaces and predict dynamic behaviors that are governed by deterministic forces due to capillary action and hydrodynamic effects. 
Although these assumptions can reasonably describe wetting phenomena in macroscale systems, random thermal fluctuations and the microscopic details of the surface must be properly considered to understand interfacial transport processes at micro- and nanoscales.    
With the advent of micro- and nanofabrication techniques a comprehensive understanding of dynamic wetting has become essential to improve traditional industrial processes such as surface coating and spraying and to fully exploit the potential of modern fabrication techniques such as micro/nanolitography and additive manufacturing (or 3D printing).
%

%
%
As the system dimensions shrink to micrometer scales and below, roughness and chemical heterogeneities inherent to natural and artificial surfaces pose a major challenge in modeling wetting processes  \cite{de1985wetting,quere2008wetting}.
Given the multiscale nature of the microscopic structure of solid surfaces it is not always feasible to define a single characteristic dimension.
Nevertheless, surface roughness and heterogeneities are usually characterized by a ``defect'' size $s_d$, determined by some relevant dimension given by the root-mean-square (rms) roughness, height autocorrelation length, or other topographic parameters.
For ``macroscopic'' defect sizes $s_d > 100$~nm, thermal fluctuations can be neglected and for low Capillary numbers the dominant forces are due to elastic deformation of the interface and pinning at localized defects  \cite{joanny1984model,robbins1987,prevost2002dynamics,de1985wetting}.
These elastic and pinning forces are merely the consequence of changes in interfacial energies as the contact line moves over random surface heterogeneities of physical and/or chemical nature.
When multiple ``macroscopic'' defects collectively distort and pin the contact line, the energy barriers preventing net displacement give rise to contact angle hysteresis  \cite{johnson1964contact,huh1977effects,oliver1980experimental,extrand1997experimental,ramos2003wetting}.    
The conventional approach to consider the effects of random surface defects with macroscopic ($s_d > 100$~nm) or mesoscopic ($s_d \simeq$~10--100~nm) dimensions consists in employing receding and advancing contact angles that are different from the Young contact angle $\theta_Y$, which is determined by minimization of energy on a perfectly smooth surface. 
Despite available predictive models based on the Wenzel  \cite{wenzel1936resistance} and Cassie-Baxter  \cite{cassie1944wettability} equations, no analytical approach has been established to quantitatively predict the degree of contact angle hysteresis from topographic parameters characterizing the surface  \cite{mchale2007cassie,marmur2009wenzel,quere2008wetting,ramiasa2014influence}.
As result, receding and advancing contact angles for static and dynamic conditions for different surfaces and liquid pairs must often be determined empirically.

%
%
It is necessary to model the effects of random thermal motion when surface defects have dimensions smaller than 100 nm and become comparable to the nanoscale thermal fluctuations of the liquid interface. 
The interplay between thermal motion and nanoscale surface features can lead to nontrivial wetting processes that are induced by thermal fluctuations of the contact line  \cite{cherry1969,marmur1994thermodynamic,rolley2007dynamics,prevost1999thermally,davidovitch2005,restagno2000thermally, ramiasa2013contact,colosqui2013,rahmani2015,colosqui2015prl}.
A few different approaches have been proposed to model the effect thermal motion and nanoscale surface defects $s_d \le 1$~nm have on the dynamics of wetting.
In the so-called molecular kinetic theory (MKT) proposed by Blake and coworkers  \cite{Blake1969,semal1999influence,de1999droplet,blake2002influence}, the effect of atomistic and nanoscale surface defects is modeled as a frictional force that dissipates the work required for the molecules in the contact line to ``hop'' over energy barriers $\Delta E$ between adsorption sites separated by a distance $\lambda \simeq s_d$.  
The virtual frictional force proposed in MKT scales linearly with viscosity and its magnitude is often comparable to hydrodynamic forces, which can make it difficult to distinguish between damping due to pinning at nanoscale defects or hydrodynamic effects  \cite{bonn2009wetting,ramiasa2014influence,de1999droplet,duvivier2011experimental}.  
Energy barriers $\Delta E=W_a$ in MKT are determined by the ``work of adhesion'' $W_a \simeq \gamma A_d (1+\cos\theta_Y)$ at localized sites, here $\gamma$ is the liquid-vapor surface tension $\gamma$ and $A_d\sim s_d^2$ the area of the adsorption site.
Predictions from MKT show agreement with experimentally observed displacement rates for different liquid pairs by assuming nanoscale defect sizes $s_d=$~0.2--1~nm (e.g., see Ref. \cite{ramiasa2014influence}).  
For consistency with the model assumptions of MKT the defect size must be smaller than 1 nm ($A_d\sim 1$~nm$^2$), which yields energy barriers $\Delta E \lesssim 10~k_B T$ (here $k_B$ is the Boltzmann constant and $T$ the system temperature).

Notably, a series of recent experimental studies on diverse systems indicate that even larger defect sizes of the order of 10 nm can induce wetting processes that are thermally activated.
For example, experimental observations report that single colloidal particles at water-oil interfaces exhibit surprisingly slow adsorption rates with time scales to reach equilibrium conditions on the order of several hours or even days  \cite{kaz2012,wang2013}.
According to conventional wetting models for perfectly spherical particles  \cite{pieranski1980,binks2006colloidal}, the adsorption dynamics of single particles is a fast monotonic decay to stable equilibrium conditions where the system energy is a global minimum.
The slow adsorption rates observed for diverse microparticles were attributed to thermally activated processes induced by surface defects with sizes ranging from 1 to 5 nm  \cite{kaz2012}.
Studies of the spreading dynamics of low viscosity liquids on surfaces with defect sizes of 10 nm report that the contact line displacement is governed by thermally activated processes  \cite{prevost1999thermally,rolley2007dynamics,rolley2009prewetting,davitt2013thermally,du2014thermally}.
These studies  \cite{davitt2013thermally,du2014thermally} indicate that energy barriers prescribing the displacement rate of the contact line are significantly smaller than the work of adhesion, and thus energy barriers $\Delta E \ll \gamma s_d^2$ induced by mesoscopic defects are smaller than predicted from the defect size.
%

%
The ``kinetics'' of contact line displacement on surfaces with mesoscopic defects $s_d =$~1--100 nm can be described by wetting models based on Kramers theory of thermally activated transitions  \cite{blake2011dynamics,colosqui2013,razavi2014}. 
%
%
In this approach, the energy barrier $\Delta E$ and separation distance $\lambda$ between long-lived metastable states can have a nontrivial relation with the defect size $s_d$ since these quantities are determined by projecting the multidimensional energy landscape parametrized by molecular positions and velocities onto a one-dimensional energy profile along the ``reaction'' coordinate describing the contact line displacement  \cite{colosqui2013,razavi2014,colosqui2015prl}.  
Theoretical models recently proposed by Colosqui {\it et al.}  \cite{colosqui2013} support the idea that kinetic rates determined via Kramers theory  \cite{kramers,hanggi1986} can predict the displacement rates of contact lines in the presence of mesoscopic defects ($s_d=$~1--10~nm).
According to these models  \cite{colosqui2013} it is possible to observe both a fast dynamic regime, governed by capillary forces and hydrodynamic friction, or a much slower kinetic regime governed by thermally activated processes.
The distance from equilibrium at which the regime crossover takes place is determined by the energy barrier magnitude and defect size, as well as the length of the contact line perimeter \cite{colosqui2013}.

Previous studies by Wexler {\it et al.} \cite{wexler2015,jacobi2015} have reported the shear-driven drainage of oil-infused microgrooves and identified conditions where a finite volume of oil is retained for indefinitely long time.
The observed steady states were analytically predicted by establishing a balance between capillary forces and the applied shear stress  \cite{wexler2015}.
The drainage dynamics far from equilibrium was approximately described by a Lucas-Washburn-type equation where thermal motion is neglected and the microgroove surfaces are assumed to be macroscopically smooth but having a receding contact angle significantly different from the Young contact angle. 
Given that the drainage of the microgrooves involves the displacement of a contact line perimeter of microscale dimensions, similar phenomena observed in the adsorption of microparticles at water-oil interfaces  \cite{kaz2012,wang2013} is expected to affect the drainage dynamics.  
Indeed, experimental observations by Wexler {\it et al.} show that the drainage dynamics close to steady-state conditions presents deviations from analytical predictions from the proposed Lucas-Washburn-type equation  \cite{wexler2015}.

In the present work we extend the Lucas-Washburn-type equation for shear-driven drainage in order to consider thermal motion and the presence of nanoscale surface roughness, by following the approach proposed by Colosqui {\it et al.} for microparticle adsorption \cite{colosqui2013}.
Atomic force microscopy (AFM) is employed to characterize the surface roughness and thus determine the defect dimensions used in the proposed wetting model for thermally activated wetting.
While the rms roughness seems to determine the magnitude of the energy barriers $\Delta E$, the height autocorrelation length appears to determine the separation distance $\lambda$ between metastable states.   
The proposed model employing mesoscopic defect sizes (3--30~nm) determined via AFM describes the drainage dynamics observed close to equilibrium conditions for different oil viscosities and applied shear rates.
The agreement between the observed contact line displacements and analytical predictions indicate that the drainage close to equilibrium is dominated by thermally activated transitions between metastable states.
Moreover, we propose a criterion for estimating the crossover point where the drainage transitions from dynamics governed by capillary and hydrodynamic forces to a kinetic regime dominated by thermally activated processes.
%

%
%
%

%
%
%
%
\section{System description \label{sec:description}}
\begin{figure}[h!]
\center
\includegraphics[angle=0,width=0.8\linewidth]{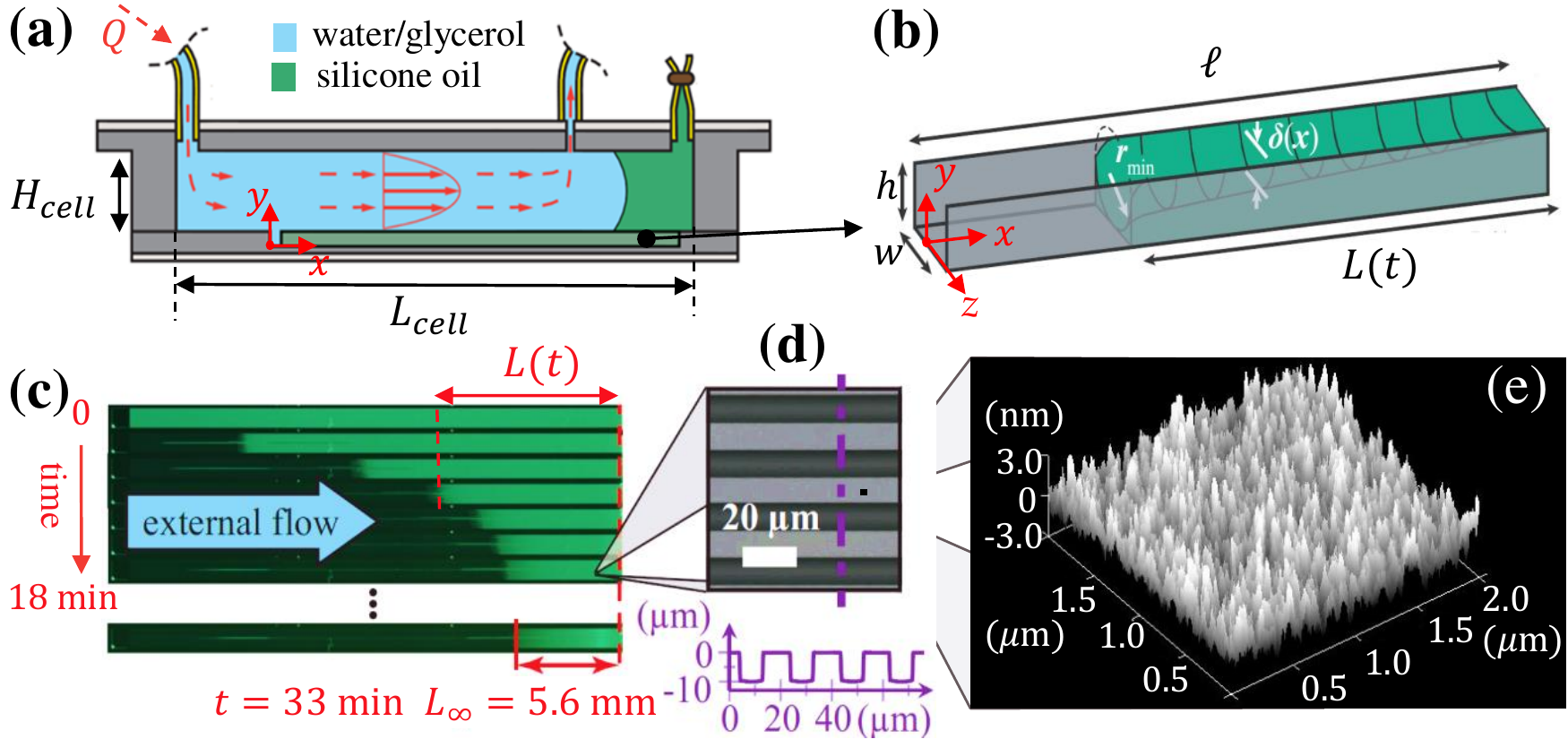}
\vskip -5pt
\caption{Experimental configuration.
(a) Schematic of the microfluidic flow cell (not to scale).  
%
%
%
An array of 50 microgrooves (bottom wall) is infused with silicone oil (green) and connected to an oil reservoir at the flow cell terminus.
%
%
(b) Schematic of the geometry of a single groove.
%
%
(c) Image sequence (3 min between images) of a sample shear-driven drainage experiment ($Q=2$~mL/min, $\mu_o=42.7$~mPa-s).
%
%
(d) Micrograph of the silicon wafer micropattern used to mold the grooves and cross-section profile and dimensions. 
Grooves appear dark gray and walls appear light gray.
(e) Topographic AFM image showing the nanoscale roughness of a sample section (2~$\mu$m~$\times$~2~$\mu$m) of the groove surface. 
}
\label{fig:1}
\vskip -10pt
\end{figure}

The experimental system consists of a rectangular microfluidic cell fabricated from Norland epoxy and sealed with a transparent glass lid for visualization purposes (see Fig.~\ref{fig:1}(a)).
The microfluidic flow cell has width $W_{cell}=7$~mm, height $H_{cell}=0.18$~mm, and length $L_{cell}= 45$~mm and is filled with a 1:1 weight mixture of glycerol and water (i.e., the outer aqueous phase) with viscosity $\mu_{aq}=5.4$~mPa~s and density $\rho_{aq}=1150$~kg/m$^3$.
There is one additional port that is 10 mm downstream of the outlet slot; this port is used for filling the oil at the beginning of the experiment, and is closed when the experiment is performed.
A syringe pump maintains constant volumetric flow rates ($Q=$1--2~mL/min) in the aqueous phase via injection of fluid through an inlet port upstream of the microgrooves. 

As illustrated in Figs.~\ref{fig:1}(a)--(b), on one wall of the microfluidic cell there is a parallel array of 50 rectangular microgrooves of width $w=9$~$\mu$m, height $h=10$~$\mu$m, and length $\ell=36$~mm, which are infused with a silicone oil that is immiscible with the aqueous phase. 
Two different silicone oils are used to infuse the microgrooves: 
1) 1,1,5,5-Tetraphenyl-1,3,3,5-tetramethyltrisiloxane
(Gelest PDM-7040), with viscosity $\mu_o=42.7$~mPa-s, density $\rho= 1061~$kg/m$^3$, and interfacial tension
(with the aqueous solution) $\gamma= 29$~mN/m; and 
2) 1,1,3,5,5-Pentaphenyl-1,3,5-trimethyltrisiloxane
(Gelest PDM-7050) with viscosity $\mu_o=201$~mPa-s, density $\rho= 1092$ kg/m$^3$, and interfacial tension
(with the aqueous solution) $\gamma= 28.2$~mN/m.
The silicone oils are mixed with Tracer Products TP-4300 UV Fluorescent Dye (cf. Fig.~\ref{fig:1}(c)) in a volume ratio of 500:1 to visualize the evolution of the dewetting process. 
The system temperature in all cases is $T\simeq 24\pm 1^{\circ}\mathrm{C}$.

%
%

After the syringe pump starts to inject the water/glycerol mixture, a finite time $t_S$ must elapse before reaching steady flow conditions with the prescribed volumetric rate $Q$.
A time $t_S=\rho_{aq} l^2/\mu_{aq}\simeq 150$~s can be estimated by considering solely diffusive effects; this time is in good agreement with experimental observations for all the flow rates studied in this work.
As shown in the image sequence in Fig.~\ref{fig:1}(c), the outer flow drives the gradual dewetting of the oil infused in the microgrooves until reaching a final finite length $L_\infty$, after which the microgrooves remain partially filled indefinitely;
the time to reach the final length $L_\infty$ is on the order of thousands of seconds under the studied conditions.
Assuming plane Poiseuille flow and a large viscosity ratio $\mu_{o}/\mu_{aq}\gg 1$, and given that the microgrooves are aligned with the outer flow, the shear stress applied at the oil-water interface is estimated as $\tau_{xy}=6\mu_{aq}Q/W_{cell}H_{cell}^2$. 
The predicted stress $\tau_{xy}$ is employed to describe experimental observations except for the case of low viscosity oil and high flow rate where the shear stress employed is 15\% smaller than analytically estimated; this deviation is attributed to the finite viscosity ratio ($\mu_o/\mu_{aq}=7.9$) for the latter case.  
The Reynolds number in the aqueous phase is $Re=(3/2) \rho_{aq} Q/ W_{cell} \mu_{aq}\simeq$~0.8--1.5, therefore small corrections (5--10\%) to the predicted stress $\tau_{xy}$ can be attributed to deviations from plane Poiseuille flow and end effects.
Since the Reynolds number in the oil phase is ${\cal O}(10^{-2})$ and the Bond number is ${\cal O}(10^{-4})$, inertial and gravitational effects can be neglected inside the microgrooves.
%

%
%
The microfluidic device is molded from Norland Optical Adhesive (NOA 81) using the ``sticker'' technique  \cite{bartolo2008,wexler2015}.
The array of microgrooves is molded from PDMS that is in turn molded from an etched silicon wafer with the nominal cross-section profile shown in Fig.~\ref{fig:1}(d).
The cross-section profile of the microgroove array presents micron-scale deviations from the nominal geometry that are below 5\% and can be observed by optical microscopy.
This small ``error of form'' is expected to cause small deviations from the flow conditions predicted for the nominal microgroove geometry (see Fig.~\ref{fig:1}(d)). 
Analysis of the microgroove surfaces is performed with a scanning probe microscope (Bruker Dimension Icon) operating in AFM tapping mode (PeakForce Tapping$^\circledR$) with a height resolution of 0.1 nm and lateral spatial resolution of 2 nm.
Topographic imaging via AFM (see Fig.~\ref{fig:1}(e)) reveals a complex random topography with nanoscale physical features resembling peaks and valleys with maximum heights and depths on the order of 3 nm and lateral dimensions reaching up to 50 nm.
As discussed in detail in the next section, the presence of nanoscale roughness is expected to cause pinning of the contact line and thermally activated processes that lead to significant deviations from the dewetting dynamics predicted for a perfectly smooth surface.   

\section{Theoretical modeling}
%
As in previous work by Wexler {\it et al.}  \cite{wexler2015}, we begin by assuming unidirectional creeping flow in the oil inside the microgrooves so that the streamwise fluid velocity $u(y,z,t)$ satisfies the governing equations $\partial u/\partial x=0$ and $\mu_o \nabla^2 u-dp/dx=0$ for mass and linear momentum balances; here, $\mu_o$ is the dynamic viscosity of the oil and $p(x,t)$ is the pressure in the oil phase.   
For the studied experimental configuration and given that the oil is much more viscous than the aqueous solution we will assume a constant pressure $p_o$ in the external aqueous phase.  
Under the assumed incompressible flow conditions the pressure inside the microgroove must vary linearly ($dp/dx=~$const.) and so must the curvature of the top free surface $\kappa=1/r(x)$ since a pressure drop $\Delta p=-\gamma/r(x)$ (for $r\ll L$) is induced by capillary effects.  
Hence, the pressure inside the oil is $p(x,t)=p_0 + (\gamma/r_{min})(x/L)$ where  \cite{wexler2015} 
\begin{equation}
r_{min}=
\left\{\begin{matrix}
w/(2\cos\theta) & \mathrm{for} & \frac{w}{h} \le 2(\mathrm{sec}~\theta+\tan\theta)\\\\ 
\frac{h}{2}\left(1+ (w/2h)^2 \right) & \mathrm{for} & \frac{w}{h} > 2(\mathrm{sec}~\theta+\tan\theta),
\end{matrix}\right.
\label{eq:rmin}
\end{equation}
is the minimum radius of curvature at the downstream end ($x=\ell-L(t)$) determined by the receding contact angle $\theta$ (see Fig.~\ref{fig:1}(b)).
A receding contact angle $\theta=56\pm 4^\circ$ has been previously determined from experimental measurements \cite{wexler2015} and since $w/h=0.9$ we have $r_{min}=w/(2\cos\theta)$ according to Eq.~(\ref{eq:rmin}).
For the assumed curvature profile of the oil-water interface the oil volume inside the microgroove is $V(t)=c_d w h L(t)$ where \cite{wexler2015} 
\begin{eqnarray}\label{eq:cd}
c_d=1-\frac{r_{min}}{h} \left(1-\sqrt{\frac{1}{4}-\frac{w^2}{16 r_{min}^2}}\right)
+ \frac{r_{min}^2}{wh} \mathrm{arcsin}\left( \frac{w}{2r_{min}}\right).
\end{eqnarray}

Conservation of mass determines that the rate of change of oil volume 
\begin{equation}
c_d w h \frac{dL}{dt}=-(q_s+q_p)
\label{eq:volume_cons}
\end{equation}
inside the grooves is determined by the volumetric flow rates $q_s$ driven by the applied shear force $F_s=\tau_{xy} w L$, and $q_p$ induced by the force $F_p=-(\gamma/r_{min}) w h$ due to capillary pressure.
Assuming creeping flow conditions and a rectangular cross-section for the liquid-filled region, analytical solution of the momentum conservation equations gives the corresponding volumetric rates and conductivities: 
\begin{equation}
q_s= \frac{c_s h^2}{\mu_o L} F_s~~\mathrm{with}~~c_s=\frac{1}{2}-\frac{4h}{w}\sum_{n=0}^\infty \frac{(-1)^n}{b_n^4}\tanh\left( \frac{b_n w}{2h}\right), 
\label{eq:qs}
\end{equation}
and
\begin{equation}
q_p=\frac{c_p h^2}{\mu_o L} F_p~~\mathrm{with}~~c_p=\frac{1}{3}-\frac{4h}{w}\sum_{n=0}^\infty \frac{(-1)^n}{b_n^5}\tanh\left(\frac{b_n w}{2h}\right).
\label{eq:qp}
\end{equation}
Here, $b_n=(n+1/2)\pi$ are the eigenvalues for each Fourier mode in the analytical solution of the momentum equation.
For the nominal microgroove height and width in the experiments of Wexler {\it et al.} \cite{wexler2015} we have $c_d=0.96$, $c_s=6.34\times 10^{-2}$, and $c_p=4.84\times 10^{-2}$.
Combining volume and momentum conservation laws embodied in Eqs.~(\ref{eq:volume_cons})--(\ref{eq:qp}) we arrive to a Lucas-Washburn-type (L-W) equation  \cite{wexler2015}  
\begin{equation}
\frac{dL}{dt}= -\frac{1}{c_d \mu_o} \left(c_s \tau_{xy} h - \frac{c_p\gamma h^2}{r_{min} L}  \right).
\label{eq:LW}
\end{equation}
This equation was derived in prior work by Wexler {\it et al.}  \cite{wexler2015} and predicts that for $t\to\infty$, for which $dL/dt=0$, the system reaches a stationary or final length
$L_{\infty}=(c_p h \gamma)/(c_s r_{min} \tau_{xy})$.
Introducing the final length in Eq.~(\ref{eq:LW}) the equation for the displacement rate takes the simple form
$dL/dt= - U_{LW} (1 -L_{\infty}/{L})$, where $U_{LW}=(c_s/c_d)(\tau_{xy} h/\mu)$ determines the maximum displacement rate attained for $L/L_{\infty}\gg 1$.
Integrating the displacement rate $dL/dt$ in Eq.~(\ref{eq:LW}) leads to an implicit expression for the column length:
\begin{equation}
t=t_S+\frac{L_\infty}{U_{LW}}
\left[\log\left(\frac{L(t)-L_\infty}{L(t_S)-L_\infty}\right)+\frac{L(t_S)-L(t)}{L_\infty}\right],
\label{eq:LWL}
\end{equation}
where $t_S$ is the time after which stationary flow conditions are attained in the aqueous phase.

A few comments are in order about the derivation of Eqs.~(\ref{eq:LW})--(\ref{eq:LWL}). 
Predictions from Eqs.~(\ref{eq:LW})--(\ref{eq:LWL}) are valid for a constant shear stress $\tau_{xy}$ assuming Poiseuille flow in the aqueous phase, and thus $t_S\simeq 150$~s in Eq.~(\ref{eq:LWL}) is the finite time required to reach steady state conditions in the outer phase (as discussed in Sec.~\ref{sec:description}).
%
%
%
%
The derivation assumes a contact line perimeter of length $s=2h+w$ that is uniform and has a constant receding contact angle $\theta$, which implies the assumption of a perfectly flat surface with constant and spatially homogeneous contact angle hysteresis.   
Nanoscale surface roughness and/or chemical heterogeneities induce spatial fluctuations of the contact line position and local contact angle that are associated with ``pinning'' at localized surface defects.
Thermally activated depinning becomes the dominant mechanism inducing contact line displacement as the system approaches the equilibrium length $L \to L_\infty$ where the effective driving force $F_d=-c_s F_s+c_p F_p \to 0$ in Eq.~(\ref{eq:LW}) vanishes. 
In the following section we proposed an extension of the L-W approach in Eqs.~(\ref{eq:LW})--(\ref{eq:LWL}) that considers the interplay between nanoscale surface defects and thermal motion so as to better characterize the drainage dynamics near equilibrium.
%

\subsection{Surface heterogeneities and thermal motion}
\begin{figure}[h!]
\center
\includegraphics[angle=0,width=0.8\linewidth]{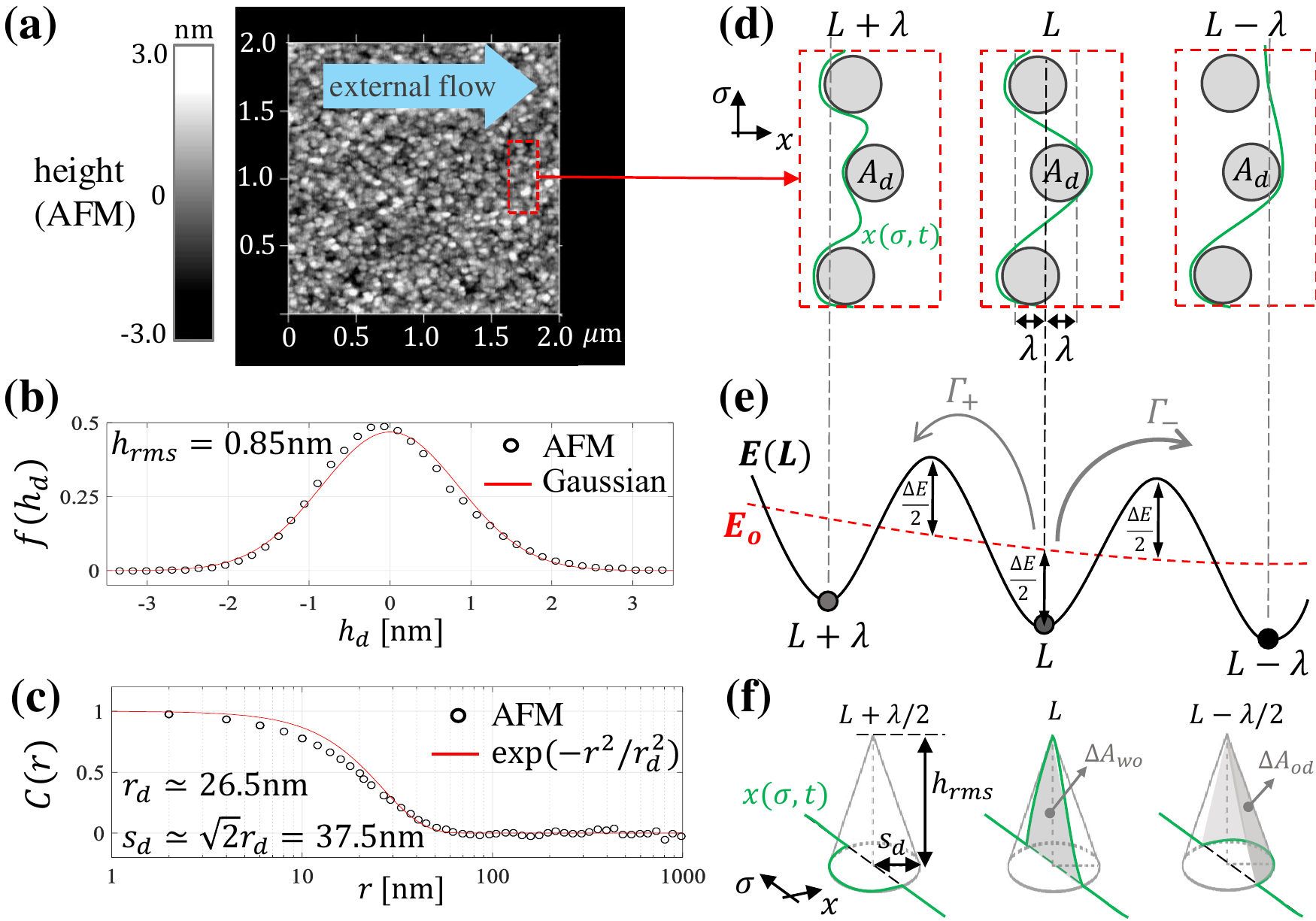}
\vskip -5pt
\caption{Nanoscale roughness and energy barriers.
(a) Two-dimensional AFM image of a sample section of the microgroove surface.
(b) Local defect height $h_d$ distribution computed from AFM data, showing a nearly Gaussian distribution ($h_{rms}=0.85$~nm).
(c) Autocorrelation function computed from AFM data (radial correlation length $r_d=26.5$~nm, defect size $s_d\simeq 37.5$~nm).  
(d) Hypothesized contact line motion induced by nanoscale defects with a projected area $A_d \simeq \pi s_d^2$. 
(e) Energy profiles $E_o(L)$ for $h_{rms}=0$ (dashed red line) and $E(L)$ for $h_{rms}>0$ (solid line).
%
%
(f) Modeled conical defect inducing an energy barrier 
$\Delta E \simeq \gamma s_d h_{rms} |1-(\pi/2) \cos\theta|=3.4~k_B T$. 
}
\label{fig:1b}
\vskip -10pt
\end{figure}

The L-W equation (Eq.~(\ref{eq:LW})) describes a one-dimensional model of drainage dynamics characterized by a single variable $L(t)$ when considering deterministic forces due to hydrodynamic and capillary effects on a macroscopically smooth surface.
As shown in Fig.~\ref{fig:1b}(a), 2D topographical imaging via AFM of a microscale section of the surface reveals a random distribution of surface defects with a maximum (peak-to-peak) height of about 6 nm. 
Analysis of the surface topography reveals a nearly Gaussian probability distribution of defect heights $h_d$ (Fig.~\ref{fig:1b}(b)) that is commonly observed for random (non-patterned) surfaces.  
The surface height presents a small rms roughness $h_{rms}=$~0.85~nm; the height distribution skewness is 0.3 and its kurtosis is 3.3, which are very close to the values expected for a Gaussian distribution.
The height autocorrelation is isotropic and presents a nearly Gaussian decay (Fig.~\ref{fig:1b}(c)) with the radial distance $r$ and a radial correlation length $r_d=26.5$~nm; 
thus we estimate a characteristic defect size $s_d=\sqrt{2} r_d \simeq 37.5$~nm and projected defect area 
$A_d= \pi s_d^2=4.4\times 10^{-3} \mu\mathrm{m}^2$. 

As illustrated in Fig.~\ref{fig:1b}(d), we will consider that the path $x(\sigma,t)~(0\le \sigma \le s)$ defined by the local streamwise position of the contact line along its perimeter $s$ is distorted by the surface defects detected in the AFM topographic image (Fig.~\ref{fig:1b}(a)). 
The average streamwise position of the contact line 
$\bar{x}(t)=(1/s)\int_0^s x(\sigma,t) d\eta$ determines the (projected) surface area $A=(l-\bar{x})s$ wetted by the liquid and thus the liquid column length $L(t)=A/s$.
Hence, the wetting/dewetting of a single surface defect with (projected) surface area $A_d$ increases/reduces the liquid column length by an amount $\lambda=A_d/s$ (see Fig.~\ref{fig:1b}(d)).
For simplicity we assume that the arclength $s\simeq 2h+w$ of the contact line is approximately constant; assuming negligible variations of $s$ implies neglecting contributions to the system energy due to line tension \cite{marmur1997line}.
We will further consider that surface defects with a finite height $h_d\simeq h_{rms}>0$ induce spatial fluctuations of characteristic magnitude $\Delta E$ in the energy $E(L)$ required to vary the liquid column length $L$, as illustrated in Figs.~\ref{fig:1b}(e)--(f).
%
%
The energy fluctuation magnitude $\Delta E$ is determined by complex morphological changes of the liquid-liquid and liquid-solid interfaces that are induced by surface defects. 
Moreover, adsorption of water or oil molecules at mesoscopic voids created by the substrate topography and interfacial phenomena induced by steric effects are likely to cause significant variations of the local surface energies.
Given this complexity, the magnitude of the characteristic energy barrier $\Delta E$ induced by surface defects will be considered as a model parameter that can be obtained by fitting experimental observations.
Nevertheless, modeling surface defects as cones with base area $A_d=\pi s_d$ and height $h_d=h_{rms}$ determined by AFM imaging we can analytically estimate an energy barrier of magnitude 
$\Delta E \simeq \gamma s_d h_{rms}|1-(\pi/2) \cos\theta|=1.4\times 10^{-20}~\mathrm{J}=3.4~k_B T$; as illustrated in Fig.~\ref{fig:1b}(f) the motion of the contact line over a modeled defect involves changes $\Delta A_{wo}=s_d h_{rms}$ in the water-oil interfacial area and $\Delta A_{od}=(\pi/2) s_d h_{rms}$ in the surface area wetted by the oil phase.
As expected the analytically estimated energy barrier vanishes for a perfectly flat surface with $h_{rms}=0$.
%

In order to incorporate the effects of nanoscale surface defects and thermal fluctuations of the contact line we will begin by considering $L(t)$ as a generalized coordinate, or reaction coordinate, determined by the surface area wetted by the oil.
Accordingly, we can recast Eq.~(\ref{eq:LW}) as 
$dL/dt= -(1/\xi) (d E_o/d L)$ 
where $\xi=c_d \mu_o s$ is an effective resistivity and
\begin{equation}
E_o(L)= s \left[c_s \tau_{xy} h L -\frac{c_p \gamma h^2}{r_{min}} \log\left(\frac{L}{L_0}\right)\right]
\label{eq:Eo}
\end{equation}
is the energy required to change the liquid column length for the case of a smooth groove with $h_{rms}=0$ ($L_0$ is an arbitrary reference length, which results in the addition of an arbitrary constant in Eq.~(\ref{eq:Eo})).
The energy profile $E_o$ has a global minimum when the stationary length is reached and thus $dE_o/dL \to 0$ as $L\to L_\infty$. 
%
%
For analytical simplicity, the effect of heterogeneities or localized surface defects will be modeled by adding a single-mode perturbation to the smooth-surface energy $E_o$ so that the energy to vary the liquid column length is 
$E(L)=E_o(L)+(\Delta E/2)\sin(2\pi(L-L_\infty)/\lambda+\varphi)$; the arbitrary phase $\varphi=-\pi/2$ is chosen so that the global energy minimum remains at $L=L_\infty$.
Given that $\lambda \ll L$, multiple local energy minima will exist at $L_o\simeq L_\infty\pm n \lambda$ ($n$ is an integer) when the system is sufficiently close to equilibrium ($L\to L_\infty$) where $dE_o/dL\to 0$.
Therefore, for $L\to L_\infty$ the system exhibits multiple metastable configurations separated by different energy barriers $\Delta E_{\pm}=E(L_o\pm\lambda/2)-E(L_o)$ in the forward/backward ($+/-$) directions and thermal motion becomes the dominant effect inducing transitions between neighboring metastable states. 

To consider thermally activated processes, we incorporate in the L-W equation (Eq.~(\ref{eq:LW})) for the column length dynamics a stochastic thermal force $F_{th}=\sqrt{2 k_B T \xi} \eta(t)$, where $\eta(t)$ is zero-mean and unit-variance Gaussian noise; this thermal force $F_{th}$ is determined by means of the fluctuation-dissipation theorem.
Including energy fluctuations caused by surface defects and stochastic forces induced by random thermal motion in Eq.~(\ref{eq:LW}) the drainage dynamics is described by a Langevin-type equation
\begin{eqnarray}
\label{eq:Langevin}
\frac{dL}{dt}= -\frac{1}{\xi} \frac{d}{d L} 
\left[E_o + \frac{\Delta E}{2} \sin\left(\frac{2\pi}{\lambda} (L-L_\infty)-\frac{\pi}{2}\right)\right]
+\sqrt{2 D} \eta(t),
\end{eqnarray}
where $D=k_B T/\xi$ is the (long-time) diffusivity along the ``reaction coordinate'' defined by the liquid column length $L$.

\subsection{Near equilibrium dynamics}
The smooth-surface energy in Eq.~(\ref{eq:Eo}) has a global minimum at $L=L_\infty$ and can be accurately approximated by a second-order Taylor expansion $E_o(L)=\textstyle{\frac{1}{2}}(d^2E_o/dL^2)|_{L=L_\infty}\times(L-L_\infty)^2$ for $L-L_\infty<(3/2)L_{\infty}$. 
Hence for $L/L_{\infty}<5/2$ we have 
\begin{equation}
E(L)= \frac{K}{2} (L-L_\infty)^2 +\frac{\Delta E}{2} \sin\left(\frac{2\pi}{\lambda} (L-L_\infty)-\frac{\pi}{2}\right), 
\label{eq:Eo2}
\end{equation}
where
\begin{equation}
K \equiv \left. \frac{d^2 E_o}{d L^2} \right|_{L=L_\infty}=\frac{c_s^2 \tau_{xy}^2 r_{min} s}{c_p \gamma}. 
\label{eq:K}
\end{equation}

According to Eq.~\ref{eq:Langevin}, as $L \to L_\infty$ and $d E_o/dL \to 0$ the column length $L$ undergoes a random walk in a periodic potential with multiple minima (i.e., metastable states) located at $L_o\simeq L_\infty \pm n \lambda$. 
Near equilibrium the column length $L(t)$ will fluctuate around the local minima $L_o$ and will suddenly transition, or ``hop'', to neighboring minima if crossing over the neighboring maxima at $L_{\pm}=L_o\pm \lambda/2$  (cf. Fig.~\ref{fig:1b}(e)). 
Following Kramers theory for thermally activated transitions \cite{kramers,colosqui2013}, the forward/backward ($+/-$) transition rates (cf. Fig.~\ref{fig:1b}(e)) are given by
\begin{eqnarray}
\label{eq:rates}
\Gamma_{\pm}(L)
=\frac{1}{2\pi\xi}
\sqrt{\frac{d^2 E(L_o)}{\partial L^2} \left|\frac{d^2 E(L_\pm)}{\partial L^2}\right|} 
\exp\left[-\frac{(E(L_{\pm})-E(L_o))}{k_B T}\right]
\end{eqnarray}
for $|L-L_o|<\lambda/2$.
When ``hopping'' between metastable states at rates given by Eq.~(\ref{eq:rates}) the average drainage speed can be estimated by a rate equation $dL/dt= \lambda (\Gamma_+-\Gamma_-)$ and thus we have \cite{colosqui2013}
\begin{equation}
\frac{d L}{dt}=-U_{H} \sinh \left(\frac{L-L_\infty}{L_H}\right),
\label{eq:kramers}
\end{equation} 
where the characteristic ``hopping'' velocity is
\begin{eqnarray}
\label{eq:VH}
U_{H}=\lambda \frac{\sqrt{4(\pi/\lambda)^4{\Delta E}^2-K^2}}{2\pi\xi}
\exp\left[-\frac{(\Delta E+K \lambda^2/8)}{k_B T}\right], 
\end{eqnarray} 
and the ``hopping'' length is
\begin{equation}
L_H=\frac{2k_B T}{K\lambda}. 
\label{eq:LH}
\end{equation} 
Integration of Eq.~(\ref{eq:kramers}) leads to  
\begin{equation}
L(t)=L_\infty+L_H \mathrm{arctanh}\left[\exp\left(-\frac{U_{H}}{L_H}(t-t_o)\right) \right], 
\label{eq:Lkramers}
\end{equation} 
where $t_o$ is an initial time arising from the integration constant.
Eq.~(\ref{eq:Lkramers}) is valid for times $t \ge t_c$ where $t_c$ is the crossover time after which the drainage dynamics is dominated by thermally activated processes. 
As elaborated in the next section, one can analytically estimate a critical crossover length $L_c$ below which forces resulting from surface heterogeneities and thermal motion are larger than forces due to hydrodynamic shear and capillary pressure.
Accordingly, the initial $t_o$ in Eq.~(\ref{eq:Lkramers}) is determined to match the experimental condition $L(t_c-t_o)=L_c$, where the crossover time $t_c$ in each experiment corresponds to the time elapsed to reach the analytically estimated length $L_c$.

\subsection{Regime crossover}
Far from equilibrium conditions where the liquid column length is much larger than the equilibrium length $L(t)\gg L_\infty$, the drainage dynamics is dominated by hydrodynamic shear and capillary forces, and can thus be described with the L-W approach in Eqs.~(\ref{eq:LW})--(\ref{eq:LWL}) \cite{wexler2015}.
As mechanical equilibrium is approached $L\to L_\infty$ and $dE_o/dL \to 0$, hydrodynamic and capillary forces balance out and the drainage of the microgrooves becomes a thermally activated process described by Eqs.~(\ref{eq:kramers})--(\ref{eq:Lkramers}). 

Here, we aim to develop a criterion for predicting the crossover from shear-driven to thermally activated drainage for different geometries and physical conditions. 
For this purpose we will analytically estimate a critical column length $L_c$ below which the dynamics is dominated by random forces due to spatial fluctuations of surface energy and thermal motion.  
For overdamped systems, the frictional force is equal to the sum $\sum F$ of all other (non-frictional) forces and thus $\xi (dL/dt)=(\sum F)$.
While according to Eq.~(\ref{eq:Langevin}) the displacement rate is $dL/dt=-(1/\xi) (dE_o/dL)$ when hydrodynamic and capillary forces dominate, 
Eq.~(\ref{eq:kramers}) determines that 
$dL/dt=-U_{H} \sinh[(L-L_\infty)/L_H]$ near equilibrium conditions where surface energy fluctuations and thermal motion dominate.
Hence, there must be critical column length $L_c$ for which 
\begin{equation}
\sinh \left(\frac{L_c-L_\infty}{L_H}\right) 
= \frac{1}{\xi U_{H}}\frac{dE_o}{dL}(L_c),
\label{eq:criterion}
\end{equation}
and forces resulting from random surface energy fluctuations and thermal motion are approximately equal to the sum of hydrodynamic and capillary forces.
Once the critical length $L_c$ is obtained by solving Eq.~(\ref{eq:criterion}) one can employ Eq.~(\ref{eq:kramers}) to determine a critical displacement rate magnitude $U_c=|U_{H}\sinh[(Lc-L_\infty)/L_H]|$ below which the drainage process is thermally activated. 

It is worth remarking that the crossover between regimes is actually a gradual process and takes place over a range of lengths $L(t)\simeq L_c$.
For the sake of simplicity, however, we will assume the transition to thermally activated drainage occurs at a ``crossover'' point determined by the critical length $L_c$ implicitly defined by Eq.~(\ref{eq:criterion}).
The integration constant in Eq.~(\ref{eq:Lkramers}) will be determined to match the critical length $L(t_c)=L_c$ that is experimentally observed at a time $t=t_c$ for each studied condition, and thus $t_o=t_c+(L_H/U_{H})\log\{\tanh[ (L_c-L_\infty)/(2L_H) ]\}$.

In prior work  \cite{colosqui2013} a simple explicit expression alternative to Eq.~(\ref{eq:criterion}) was proposed to estimate the critical distance from equilibrium below which the final relaxation regime is dominated by thermally activated transitions between metastable states.
According to Eq.~(\ref{eq:Eo2}), metastable states induced by local energy minima where $dE/dL=0$ can only exist for sufficiently small column lengths 
$L \le L_\infty + (\pi \Delta E)/(K\lambda)$.
Hence, the approach to equilibrium is dominated by thermally activated transitions below a crossover length $L_c$ given by \cite{colosqui2013}
\begin{equation}
\frac{L_c-L_\infty}{L_H} =  \alpha \frac{\pi}{2} \frac{\Delta E}{k_B T}
\label{eq:Lc_approx}
\end{equation}
where $\alpha<1$ is a scaling factor smaller than unity.
As shown in the next section, the simple crossover criterion in Eq.~(\ref{eq:Lc_approx}) yields agreement with Eq.~(\ref{eq:criterion}) and experimental results for $\alpha=$~0.2--0.25.
%

%

\section{Results}
%
\begin{figure}[t]
\center
\includegraphics[angle=0,width=0.8\linewidth]{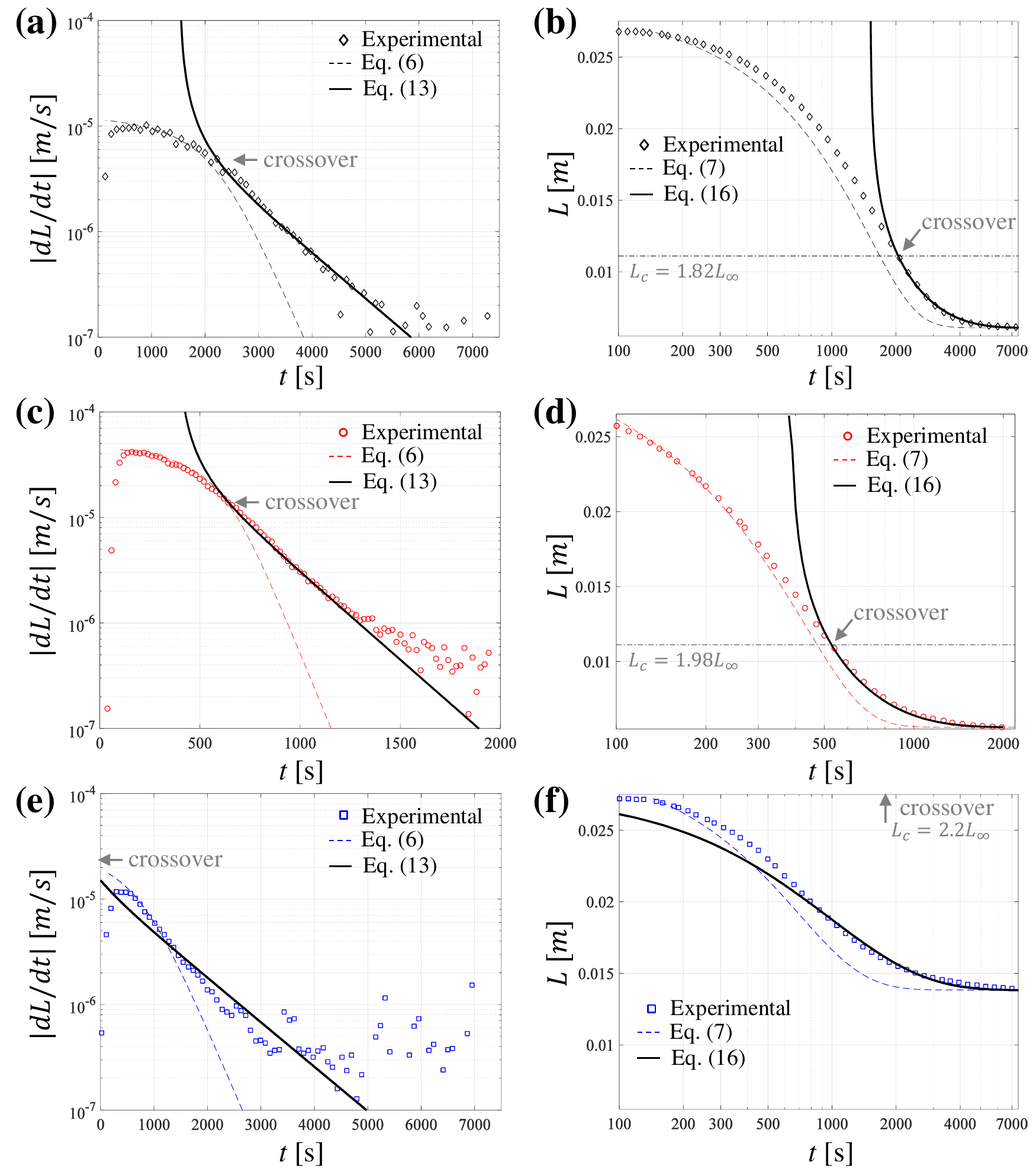}
\vskip -5pt
\caption{Displacement rate magnitude $|dL/dt|$ and column length $L(t)$ versus time for three different experimental conditions.
(a-b) Case (i): $Q=2$~mL/min and $\mu_o=201$~mPa-s. 
(c-d) Case (ii): $Q=2$~mL/min and $\mu_o=42.7$~mPa-s.
(e-f) Case (iii): for $Q=1$~mL/min and $\mu_o=42.7$~mPa-s.
Markers: experimental results.
Dashed lines: analytical predictions from Eq.~(\ref{eq:LW}) and Eq.~(\ref{eq:LWL}) adopting $t_S=150$~s. 
Solid lines: analytical predictions for drainage dominated by thermally activated processes (Eqs.~(\ref{eq:kramers})--(\ref{eq:Lkramers})) using $\lambda=0.15$~nm and $\Delta E=3.4~k_B T$ ($T=24^\circ C$).
Dashed-dotted (horizontal) lines: predictions from Eq.~(\ref{eq:criterion}) for the crossover length $L_c$.
The initial time $t_o=t_c+(L_H/U_{H})\log\{\tanh[ (L_c-L_\infty)/(2L_H)]\}$  in Eq.~(\ref{eq:Lkramers}) is determined to match the experimentally observed length at the crossover point $L(t_c-t_o)=L_c$.
}
\label{fig:2}
\end{figure}

The length of the wetted portion of a groove is determined by using automated image analysis on macroscale photographs with a pixel size of 12.5 $\mu$m.  
The pixel intensity is high in places that are wetted with oil (due to fluorescence) and low elsewhere.  
The upstream limit of the wetted length is determined by plotting the pixel intensity along the length of a groove, and finding the location where the slope changes most rapidly by applying a third-order Savitzky-Golay filter with a window size of 50--70 pixels.  
These images are taken every 10 seconds, yielding a limit to the resolvable velocity of approximately $10^{-6}$ m/s.

Three different experimental conditions are studied where the outer flow rate and viscosity of the infused oil are varied: 
(i)   $Q=2$~mL/min and $\mu_o=201$~mPa-s  (cf. Figs.~\ref{fig:2}(a)--(b)), 
(ii)  $Q=2$~mL/min and $\mu_o=42.7$~mPa-s (cf. Figs.~\ref{fig:2}(c)--(d)), and
(iii) $Q=1$~mL/min and $\mu_o=42.7$~mPa-s  (cf. Figs.~\ref{fig:2}(e)--(f)).
The displacement rate $dL/dt$ and time evolution of the column length $L(t)$ measured experimentally are compared in Fig.~\ref{fig:2} against analytical predictions from the L-W approach (Eqs.~(\ref{eq:LW})--(\ref{eq:LWL})) and the theory based on thermally activated transitions between metastable states (Eqs.~(\ref{eq:kramers})--(\ref{eq:Lkramers})).

As discussed in Sec.\ref{sec:description}, a finite time $t_S= 150$~s is employed in Eq.~(\ref{eq:LWL}) to consider the time elapsed before steady flow is attained in the aqueous phase; this is in agreement with experimental observations for the displacement rate magnitude reported in Fig.~\ref{fig:2}. 
For case (ii) where the highest volumetric rate ($Q=2$~mL/min) is employed and the liquid phase has the lowest viscosity ($\mu_o=42.7$~mPa-s), the shear stress value $\tau_{xy}=4.04$~Pa employed in Eqs.~(\ref{eq:LW})--(\ref{eq:LWL}) was 15\% lower than predicted by assuming plane Poiseuille flow and a large viscosity ratio. 
For the other experimental conditions the shear stress employed in Eqs.~(\ref{eq:LW})--(\ref{eq:LWL}) was the one predicted by assuming plane Poiseuille flow; i.e., $\tau_{xy}=4.75$~Pa for case (ii), and $\tau_{xy}=2.38$~Pa for case (iii). 
After steady flow conditions are attained for $t\ge t_S$, there is good agreement between experimental observations and analytical predictions from L-W equations (Eqs.~(\ref{eq:LW})--(\ref{eq:LWL})) during the initial stages of drainage where $L(t)<L_c$ and hydrodynamic shear and capillary forces are expected to dominate. 
%

\begin{figure}[h!!]
\center
\includegraphics[angle=0,width=0.8\linewidth]{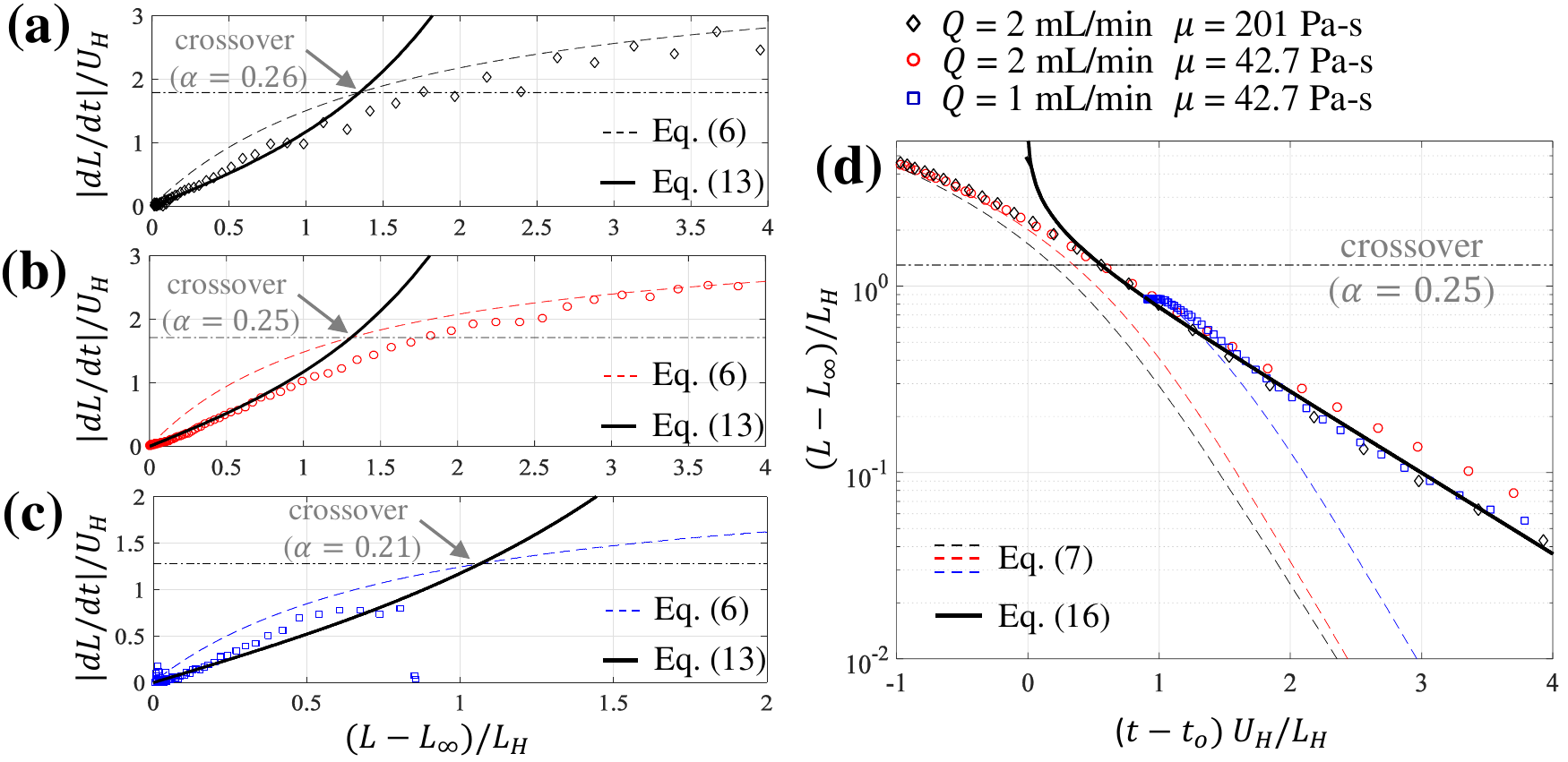}
\vskip -5pt
\caption{Approach to final equilibrium length $L_\infty$ for three different experimental conditions:
Case (i): $Q=2$~mL/min, $\mu_o=201$~mPa-s, $L_H/U_{H}=1087$~s 
($U_{H}=3.44\times 10^{-6}$~m/s, $L_H=3.74\times 10^{-3}$~m). 
Case (ii): $Q=2$~mL/min, $\mu_o=42.7$~mPa-s, $L_H/U_{H}=261.1$~s
($U_{H}=1.61\times 10^{-5}$~m/s, $L_H=4.21\times 10^{-3}$~m).
Case (iii): $Q=1$~mL/min, $\mu_o=42.7$~mPa-s, $L_H/U_{H}=1029$~s
($U_{H}=1.65\times 10^{-5}$~m/s, $L_H=1.7\times 10^{-2}$~m).
(a) Normalized displacement rate magnitude $|dL/dt|/U_{H}$ versus normalized distance from equilibrium length $(L(t)-L_\infty)/L_H$. 
(b) Distance from equilibrium $L(t)-L\infty$ versus normalized time (length shown in logarithmic scale). A nearly exponential decay with a characteristic time $T_H=L_H/U_{H}$ is observed for all studied cases.
Markers: experimental results for cases (i)--(iii).
Solid lines: analytical predictions from Eqs.~(\ref{eq:kramers})--(\ref{eq:Lkramers}) using $\lambda=0.15$~nm and $\Delta E=3.4~k_B T$ ($T=24^\circ C$).
Dashed-dotted (horizontal) line: analytical estimation for the crossover length $L_c$ from Eq.~(\ref{eq:Lc_approx}). 
}
\label{fig:3}
\end{figure}
%

As the system approaches the final equilibrium length $L_\infty$ there is a crossover to a slower drainage process predicted by Eqs.~(\ref{eq:kramers})--(\ref{eq:Lkramers}), which are valid when the dynamics are dominated by thermally activated processes.
In all studied cases, the period between metastable configurations $\lambda=\pi s_d^2/(2h+w)=0.15$~nm was determined by the defect size $s_d \simeq 37.5$~nm obtained from AFM surface imaging (cf. Fig.~\ref{fig:1b}(c)).
In order to fit experimental results reported in Figs.~\ref{fig:2}--\ref{fig:3} an energy barrier magnitude 
$\Delta E \simeq 3.4~k_B T$ ($T=24^{\circ}\mathrm{C}$) is employed for all cases.
Notably, the value of the energy barrier employed to fit experimental observations can be predicted via simple geometric arguments (cf. Fig.~\ref{fig:1b}) for the three studied conditions where the flow rate, viscosity, and surface tension are varied.
Moreover, the crossover criterion in Eq.~(\ref{eq:criterion}) (see dashed-dotted horizontal lines in Fig.~\ref{fig:2}) can be used to estimate the critical lengths $L_c$ below which the drainage becomes a thermally activated process and $L(t)$ is governed by Eq.~(\ref{eq:Lkramers}). 
For the experimental conditions in case (i) (cf. Figs.~\ref{fig:2}(a)--(b)) the crossover to thermally activated drainage occurs for $t_c\simeq 2000$~s when the column length is $L(t_c)=11.2$~mm, which is about two times larger than the expected equilibrium length (i.e., $L_c=1.9 L_\infty$).
In agreement with experimental observations for cases (ii) and (ii) (cf. Figs.(~\ref{fig:2})(c)--(f)), Eq.~(\ref{eq:criterion}) predicts an increase in the crossover length and an earlier transition to thermally activated drainage when the liquid viscosity is reduced.
In particular, the crossover criterion (Eq.~(\ref{eq:criterion})) indicates that for the lower flow rates employed in case (iii) (cf. Figs.~\ref{fig:2}(e)--(f)) the crossover length is larger than the microgroove length and the entire drainage dynamics may be thermally activated.

According to the theoretical model leading to Eqs.~(\ref{eq:kramers})--(\ref{eq:Lkramers}), all experimental observations near equilibrium conditions can be collapsed to a single curve when normalizing with the characteristic ``hopping'' velocity $U_{H}$ and length $L_H$ defined by Eq.~(\ref{eq:VH}) and Eq.~(\ref{eq:LH}), respectively.
Indeed, Figs.~\ref{fig:3}(a)--(c) report that the displacement rate magnitude closely follows the single curve predicted by Eq.~(\ref{eq:kramers}) for all studied cases (i)--(iii). 
Similarly, the distance $L(t)-L_\infty$ between the column length and the expected equilibrium length follows the single trajectory predicted by Eq.~(\ref{eq:Lkramers}) when normalized by the corresponding values of $U_{H}$ and $L_H$ for each case (Fig.~\ref{fig:3}(d)).
The linear decay in the displacement rate magnitude for $(L-L_\infty)/L_H<1$ indicates an exponential relaxation, $L(t)-L_\infty \propto \exp(-t/T_H)$, near equilibrium conditions with a relaxation time $T_H=L_H/U_{H}$ varying from about 200 to 1000 s (cf. Fig.~\ref{fig:3}(b)).  
In addition we observe that the simple crossover criterion in Eq.~(\ref{eq:Lc_approx}) can predict the crossover length $L_c$ for scaling factors $\alpha\simeq$~0.2--0.25. 
\section{Conclusions}
The analysis and experimental observations in this work indicate that the interplay between nanoscale surface roughness and thermal motion needs to be carefully considered in order to describe the dynamics of drainage and imbibition in microscale capillaries.
In the presence of significant energy barriers induced by nanoscale surface defects, the interface displacement is dominated by random thermally activated transitions between metastable states.
These random transitions give rise to a ``kinetic'' regime in the evolution of the surface area wetted by one or other phase that cannot be described by conventional (continuum-based) wetting models (e.g., L-W equations) considering solely deterministic forces due to hydrodynamic and capillary effects. 
Therefore we have proposed a stochastic Langevin equation that can be used to describe both the (far-from-equilibrium) dynamic and (near-equilibrium) kinetic regimes observed in the shear-driven drainage of microcapillaries infused with viscous liquid. 
The proposed model can be adopted to describe numerically diverse wetting processes, such as spreading of microdroplets or colloidal particle adsorption, where thermal motion and nanoscale surface roughness give rise to the same fundamental phenomena considered in this work.

To describe analytically the kinetic regime dominated by thermally-activated processes, we have employed a rate equation where transition rates are predicted by Kramers theory.
Furthermore, we have considered an energy profile exhibiting multiple metastable states with a characteristic period $\lambda=0.15$~nm and separated by a characteristic energy barrier $\Delta E \simeq 3.4~k_B T$.
In the model proposed in this work, both the period and energy barrier are determined by nanoscale defects with characteristic size $s_d\simeq 37.5$~nm and rms height $h_{rms}=0.85$~nm that are observed in AFM topographic images.
It is worth noticing that an energy barrier of magnitude $3.4~k_BT$ corresponds to the work of adhesion 
$W_a=\gamma(1+\cos\theta)A_a$ on a molecular adsorption site of area $A_a=0.32$~nm$^2$.
Thus, fitting experimental results by using an alternative wetting model such as MKT would have led us to infer that the drainage dynamics near equilibrium is caused by surface defects of molecular dimensions $s_d\simeq\sqrt{A_a}=$~0.6~nm.
Notably, AFM imaging of the studied surfaces reported the presence nano- and mesoscale defects with much larger dimensions ($s_d>10$~nm) and areas ($A_d>100\mathrm{nm}^2)$).     
The model employed in this work determines that the very small separation between metastable states ($\lambda\sim{\cal O}(10^{-10}\mathrm{m})$) is given by the ratio of the surface defect area ($A_d\sim{\cal O}(10^{-15}\mathrm{m})$) to the contact liner perimeter ($s\sim{\cal O}(10^{-5}\mathrm{m})$), i.e., it is not directly prescribed by the physical distance between surface defects.  
The proper definition of model parameters made it possible to predict both the crossover to the kinetic regime and the kinetic relaxation rate for all of the studied experimental conditions.

The analysis in this work shows that it is feasible to characterize the nanoscale surface topography, using AFM or alternative approaches, and then determine the system dimensions (e.g., capillary height and width) that will produce a desired drainage dynamics.
While the final retention length $L_\infty$ is prescribed by specific geometric and physical parameters, the time to reach the final length can be significantly  reduced/increased by (i) reducing/increasing the crossover length $L_c$ to the kinetic regime and (ii) decreasing/increasing the kinetic relaxation time $T_H=L_H/U_{H}$, which varies exponentially with the energy barrier $\Delta E$ prescribed by the surface defect area $A_d$.
The models employed in this work could aid the design of nanostructured surfaces to control the dynamics of drainage of capillaries as well as other wetting processes in microscale systems. 

We thank Dr. Chung-Chueh Chang at the Stony Brook University (SBU) ThInc for performing AFM imaging of the microgrooves samples. 
CEC acknowledges support from the SEED Grant Program by Brookhaven National Laboratory and SBU.
JSW, YL, and HAS acknowledge support from ONR MURI Grants No. N00014-12-1- 0875 and No. N00014-12-1-0962 (Program Manager Dr. Ki-Han Kim)
%

%
%

%

%

\end{document}